\documentclass[aps,prl,twocolumn]{revtex4}
\usepackage{amsmath}
\usepackage{bm}

\usepackage{color}

\usepackage{mathrsfs}
\usepackage{amsfonts}
\usepackage{graphicx}
\usepackage{dcolumn}
\usepackage[normalem]{ulem}

\begin{document}
\title{Molecular templates of spin textures on superconducting surfaces}

\author{Cristina Mier}
\affiliation{Materials Physics Center
               (UPV/EHU),
  20018 Donostia-San Sebasti\'an, Spain}
\author{Benjamin Verlhac}
\author{L\'eo Garnier}
\affiliation{Universit\'e de Strasbourg CNRS, IPCMS, UMR 7504, F-67000 Strasbourg, France}
\author{Roberto Robles}
\affiliation{Centro de F{\'{\i}}sica de Materiales
              CFM/MPC (CSIC-UPV/EHU),  20018 Donostia-San Sebasti\'an, Spain}
\author{Laurent Limot}
\affiliation{Universit\'{e} de Strasbourg, CNRS, IPCMS, UMR 7504, F-67000 Strasbourg, France}
\author{Nicol{\'a}s Lorente}
\affiliation{Centro de F{\'{\i}}sica de Materiales
              CFM/MPC (CSIC-UPV/EHU),  20018 Donostia-San Sebasti\'an, Spain}
\affiliation{Donostia International Physics Center (DIPC),  20018 Donostia-San Sebasti\'an, Spain}
\author{Deung-Jang Choi}
\email{deungjang.choi@ehu.eus}
\affiliation{Materials Physics Center
               (UPV/EHU),  20018 Donostia-San Sebasti\'an, Spain}
\affiliation{Donostia International Physics Center (DIPC),  20018 Donostia-San Sebasti\'an, Spain}

\begin{abstract}
We create ordered islands of magnetically anisotropic nickelocene molecules on a Pb
(111) substrate.  By using inelastic electron tunneling spectra (IETS) and
density functional theory, we characterize the magnetic response of these islands. 
This allows us to conclude that the
islands present local and collective magnetic excitations. Furthermore, we show 
that nickelocene islands present complex non-collinear spin patterns
on the superconducting Pb (111) surface, opening the possibility of using
molecular arrays to engineer spin textures with important implications
on topological superconductivity.
\end{abstract} 
\date{\today} 
\maketitle

%Introduction

Topological superconductors are new states of matter that are
not found in Nature.  Low-dimensionality and
chirality are key ingredients to create superconducting phases
of different topology. A particularly interesting strategy
\cite{Beenakker2011,Pientka2013,Pientka2015,Yazdani2013} is to
create arrays of atomic spins on a superconductor. Some inspiring
results have been obtained by growing arrays of Fe atoms on Pb (111)
\cite{Yazdani,Yazdani2017}, and atomically placing Fe atoms on
Re and Ta surfaces \cite{Wiesendanger2018,Jens,Wiesendanger2019}.
Experiments~\cite{Wiesendanger2019} and theory~\cite{Bernevig} show that
topological superconductors can be obtained from islands of magnetic
impurities adsorbed on an s-wave superconductor. Compelling experimental
and theoretical evidence~\cite{Wiesendanger2019} show the appearance of
helical Majorana states at the island boundaries in the dense-impurity
regime.  Furthermore, the dilute-impurity regime is also very interesting,
leading to topological phases that  can present
numerous helical Majorana states~\cite{Bernevig}.

\begin{figure}
\begin{center}
%\hspace{-1.5 cm}
  \includegraphics[width=0.48\textwidth]{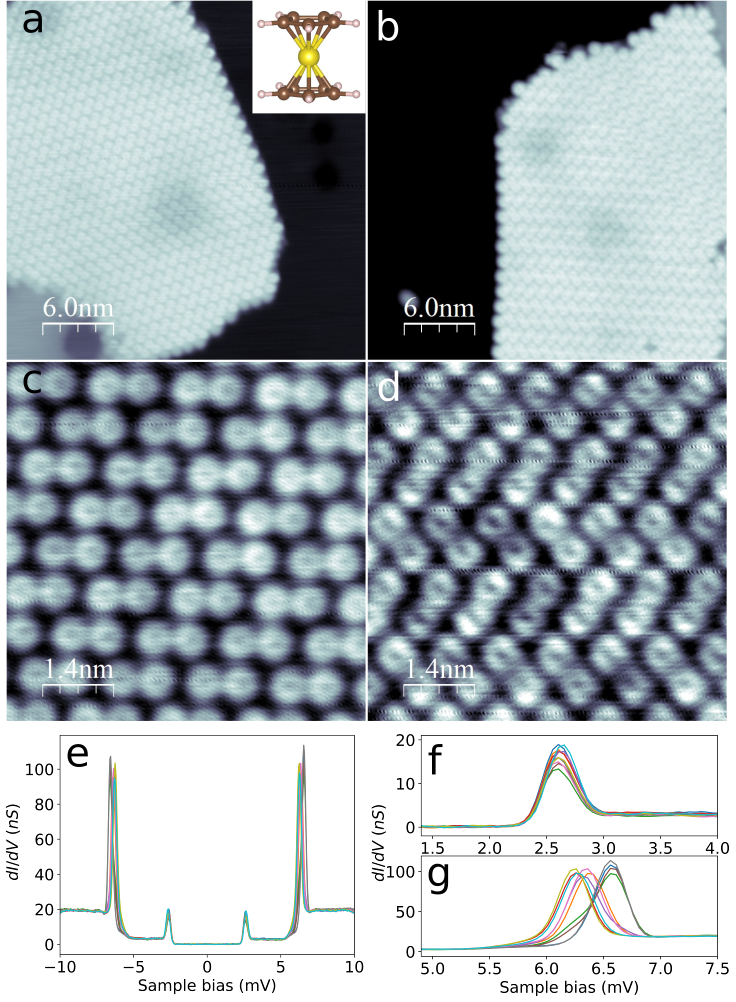}
\end{center}
\caption{(a) {Constant-current image of an} island of Nc on Pb
(111) where the Nc dimers are aligned ($-60$ mV, $20$ pA, $30\times30$
nm$^2$). Inset: the Nc molecule consists of a Ni atom (yellow) capped by C
(brown) pentagons {(or cyclopentadienyls)} with H-saturated bonds (white),
the {gas-phase} distance between cyclopentadienyls is 3.6 \AA. (b)
Constant-current image of Nc dimers arranged in zig-zag ($-80$ mV, $40$
pA, $30\times30$ nm$^2$). (c) The \textit{paired} arrangement of Nc dimers
is evidenced at \textcolor{black}{shorter tip-surface distances} ($-20$
mV, $20$ pA, $7.0\times7.0$ nm$^2$). (d) The zig-zag or \textit{compact}
arrangement is clearly seen at shorter tip-surface distances. (e) $dI/dV$
obtained with a superconducting tip over several molecules at different locations. (f)
The quasiparticle (QP) peak consistently appears at $2.61\pm0.02$ mV, but
the inelastic peak, (g), depends on the measured molecule. }
\label{figure1}
\end{figure}

The scanning tunneling microscope (STM) is an appropriate tool to study
the above systems \cite{Yazdani2018,Choi2019}. Indeed, the STM is able
to assemble atomically-precise 1D structures reaching beyond 100 spins on
superconducting Re \cite{Wiesendanger2018}.  However, this technique has
been unsuccessful on Pb surfaces \cite{Yazdani,Ruby2015} mainly due to the
large atomic mobility of Pb atoms on the surface.  A different strategy is
to use molecules that can self-assemble and create directional structures
\cite{Katharina1}.  From heavy-fermion-like lattices \cite{Tsukahara,Ho},
and quantum-phases on superconductors \cite{Katharina1,Katharina2} to
functionality \cite{Jascha}, molecules show rich spectra of physical
processes and properties.  Molecules can possess large spins and strong
magnetic anisotropies \cite{Sessoli} together with complex patterning. 
As a consequence, they can give rise to non-collinear spin textures extending
the possibilities for topological superconductivity.

We report on molecular structures formed by nickelocene
(Nc) molecules adsorbed on Pb (111) studied by low-temperature
STM. 
\textcolor{black}{The inset of Fig. \ref{figure1} (a) shows that Nc is
composed by two cyclopentadienyl rings (C$_5$H$_5$) and a Ni atoms in between. It  is a $S=1$ spin 
with sizeable hard-axis longitudinal magnetic anisotropy perpendicular to the cyclopentadienyl rings
ranging from $D=3.4$ to $4.0$ meV} depending on the substrate
\cite{Ormaza2017a,Ormaza2017b,Verlhac2019,Bachellier2020}.
Studies on copper surfaces
\cite{Ormaza2017a,Ormaza2017b,Bachellier2020,Bachellier2016} show that
vertical Nc molecules are intercalating with horizontal ones to form the
molecular assembly.
Due to the molecular magnetic anisotropy, 
such an assembly has the potential for producing a complex spin texture.
\textcolor{black}{
Here, we use the sub-meV energy resolution of a superconducting-tip 
 to investigate Nc molecular islands on Pb (111). With the help
of first-principles calculations, we find that indeed the molecular
arrangement within these islands imposes a non-collinear texture of
interacting spins. We further show that the substrate-induced
interactions lead to a dilute spin regime 
that is interesting for obtaining topological superconducting phases.}

\textcolor{black}{
We deposit Nc molecules at room temperature in ultra high
vacuum conditions on a Pb (111) surface kept at a temperature ranging from
70-150 K.}
Figure \ref{figure1} shows constant-current STM images of
a representative set of molecular islands obtained on Pb (111) and at a
temperature of 2.5 K.  We find two types of arrangements after depositing
 the molecules.  The linear arrangement of
dimers, Fig. \ref{figure1} (a) largely coincides with the \textit{paired}
arrangement found on Cu (100) and Cu (111) \cite{Bachellier2016}. The
 zig-zag configuration, Fig. \ref{figure1} (b), agrees with the
\textit{compact} pattern of the previous studies \cite{Bachellier2016}.
This can be clearly seen in the smaller-area images of Fig. \ref{figure1}
(c) and (d). Different combinations of the two arrangements can take
place in the same island, either repeating one pattern or the other
in different domains.  We perform density functional theory (DFT)
studies~\cite{SI} that show the two patterns share the same stabilization
energy, compatible with the coexistence of both phases.

The differential conductance {($dI/dV$)} measured on the above systems
gives interesting information on their superconducting and magnetic
properties. Figure \ref{figure1} (e) shows typical $dI/dV$ 
measurements on molecules of the paired phase. The STM tip
is coated by Pb giving a  superconducting gap
 $\Delta$ slightly smaller than the bulk Pb gap of
$1.35$ meV\cite{Webb}, due to finite-size effects \cite{Loth2014}. 
As a consequence, the surface superconducting gap
appears at $\Delta_{tip}+\Delta_{substrate}\approx 2 \Delta$ in the $dI/dV$ data due
to the convolution with the tip electronic 
structure~\cite{QKX,Ruby2015a,Choi2017},
confirmed by the two quasiparticle (QP) peaks appearing
at $\pm 2.61 \pm 0.02$ mV in Fig. \ref{figure1} (e).
Zooming in into these peaks, Fig. \ref{figure1} (f),
 shows that they are  a property of 
the substrate and they do not depend on the molecule where the
spectra are taken. However, the structure appearing about $6.4$ mV
strongly depends on the target molecule, Fig. \ref{figure1} (g). 
\textcolor{black}{These peaks correspond to the inelastic electron
tunneling spectra (IETS) of the Nc magnetic excitation on a superconductor~\cite{Liljeroth2019}.}
Indeed, these peaks match the excitation energy of the molecular spin from the $S_z=0$ ground
state to the $S_z=\pm 1$ excited states \cite{Ormaza2017a,Ormaza2017b,Bachellier2020,Verlhac2019}
with excitation energy $D=6.4 \pm 0.2 -2.6=3.8 \pm 0.2$ meV
shifted by $2 \Delta$ due to the presence of
the tip and substrate superconducting gaps~\cite{Liljeroth2019}.

\begin{figure}
\begin{center}
  \includegraphics[width=0.48\textwidth]{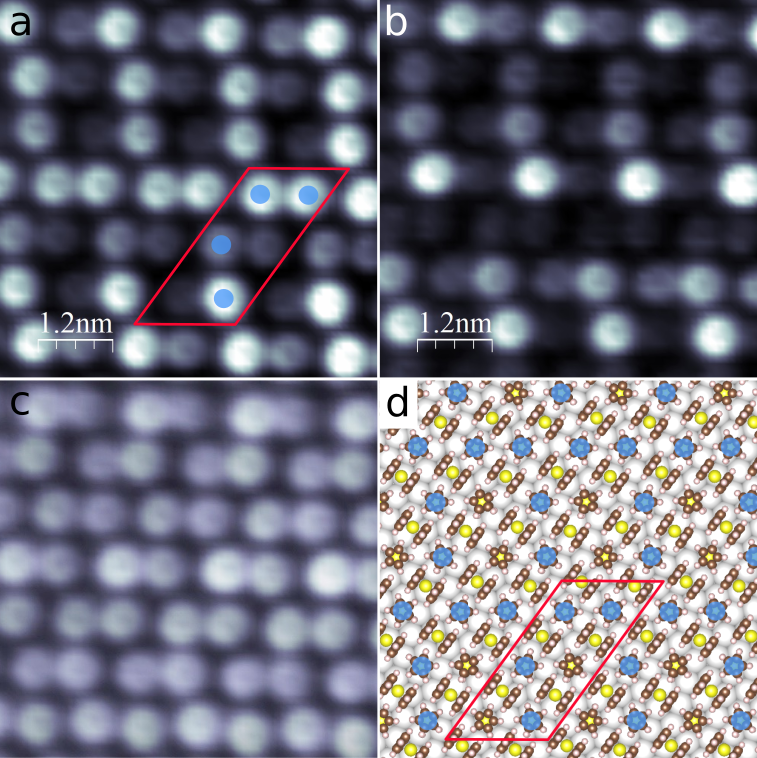}
\end{center}
  \caption{
(a) Constant-height $dI/dV$ map taken at 6.6 mV,
and (b) at 6.0 mV over the \textit{paired} phase, Fig.~\ref{figure1} (c).
(c) The two maps are merged into one single map, showing the same dimer
structure as the constant-current image and hence their complementarity.
(d) Atomic structure of the unit cell for the
aligned-dimer phase. The maps in (a) and (b) point at a unit cell formed by
three vertical and three horizontal dimers aligned
within an angle of 50$^\circ$ with the $[\bar{1}01]$ direction of the Pb (111) surface.
The Nc molecules with the smallest computed
MAE (FCC and bridge sites) are marked by a circle on the corresponding vertical molecule. 
The emerging pattern perfectly matches the $dI/dV$ map at 6.6 mV in (a). The complementary
pattern is also obtained and it is due to the molecules on the remaining top sites that present a 
lower MAE than on the FCC and bridge sites.
\label{figure2}}
\end{figure}

Figure~\ref{figure2} shows two constant-height $dI/dV$ maps of
Figure~\ref{figure1} (c).  Panel (a) presents the spatial distribution
of the $dI/dV$ at $V=6.6$ and (b) at $V=6.0$ mV.  Figure \ref{figure2}
(c) shows the two merged $dI/dV$ maps  reproducing the dimer pattern
of Fig. \ref{figure1} (c).  Both maps, Fig. \ref{figure2} (a) and (b),
show a quasi-periodic pattern involving three dimers, instead of the
single-dimer periodic structure of Fig. \ref{figure1} (c).  In order
to understand these spatial IETS peak patterns, Fig.\ref{figure2}
(a) and (b), the location of the dimers on the substrate needs to
be found.  To do this, we perform STM images on the edge of the
compact island at different tunneling currents to obtain atomic
resolution of the Pb (111) substrate \cite{SI}.  From these data,
the unit cell of Fig. \ref{figure2} (d) emerges. As in previous studies
\cite{Ormaza2015,Bachellier2016,Bachellier2020} a combination of vertical
(with the molecular axis perpendicular to the surface)
and horizontal (parallel to the surface) Nc has to be considered.  The new pattern consists of
two vertical molecules and two horizontal molecules.
The experimental
angle of the structure is $ 50^\circ\pm 4^\circ$, in good agreement with
 the angle of $53.5^\circ$ we find by using DFT to optimize a structure
based on 3 vertical and 3 horizontal dimers \cite{SI}.

We further compute different molecular-layer energies by
shifting the molecular structure over the surface.
All the computed structures are basically
equivalent, with the larger energy difference being only $20$ meV per molecule.
This \textcolor{black}{feeble} dependence on the actual adsorption sites show the small corrugation of the substrate-molecule
interaction. As a consequence, the molecular structure presents
some degree of incommensurability~\cite{c60} that translates in the quasi-periodicity
of the IETS pattern Fig. \ref{figure2} (a) and (b).
The two types of molecules (vertical and horizontal) lead to different interaction with the substrate. 
The horizontal molecules are weakly interacting via vdW forces \textcolor{black}{and their computed adsorption
energy is $0.5$ eV/molecule.}  However, the vertical molecules do present 
hybridization between their frontier $\pi$ orbitals and the $sp$-electronic structure
of the Pb (111) surface \textcolor{black}{with $0.7$ eV of adsorption energy per molecule}.

\begin{figure}
\begin{center}
  \includegraphics[width=0.40\textwidth]{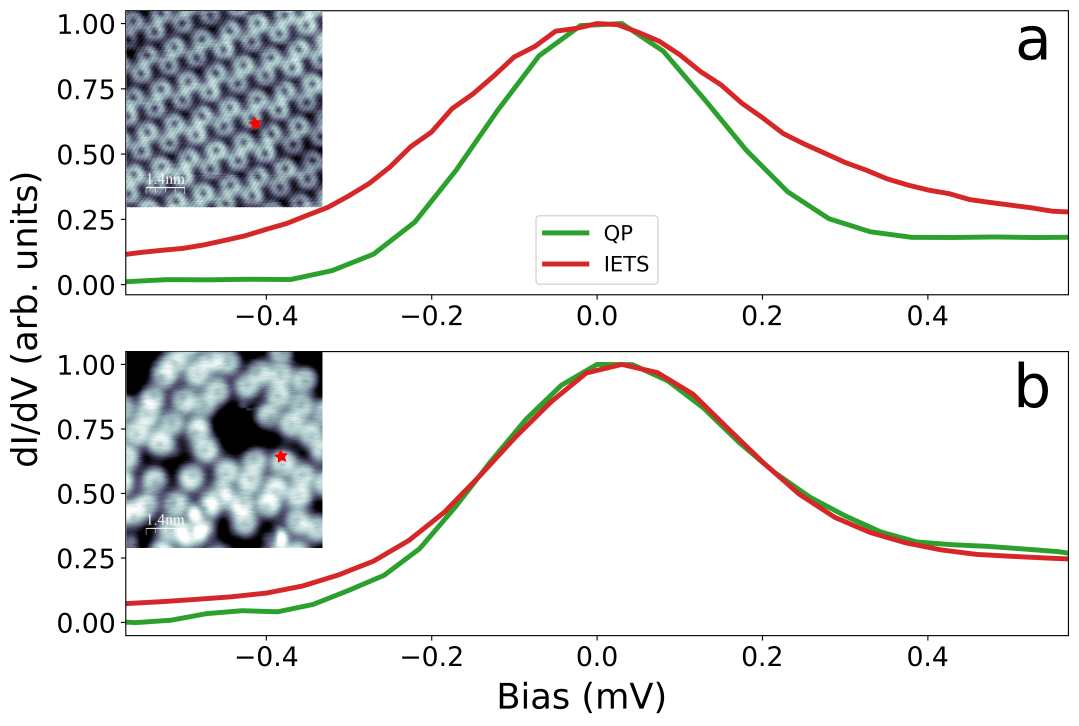}
\end{center}
  \caption{
(a) Lineshape of the IETS peak superimposed on the QP peak by shifting
their respective bias origins and rescaling the peaks, for (a) a molecule in an order island
(inset shows the tip location) and (b) for a molecule with few neighbors
in a disordered island (see inset). Molecules with large regular
neighbor coordination show an extra source of broadening for the
IETS peak.  
} 
\label{figure3} 
\end{figure}

Extra information of the magnetic interactions at play can be obtained
from the IETS lineshape.  Figure~\ref{figure3} superimposes the $dI/dV$
lineshape of the IETS on the QP peaks. In the case of (a), the $dI/dV$
spectra were obtained on a highly-coordinated molecule in a ordered island
\textcolor{black}{shown in the inset}. The case of (b) corresponds to
a lowly-coordinated molecule in a disordered island\textcolor{black}{,
see inset. These figures hint to}
 an extra source of broadening for the IETS peak in the ordered case.

In order to {explain} the magnetic IETS
data, we perform non-collinear DFT calculations with spin-orbit
coupling.
We obtain a clear relation between the
 magnetic anisotropy energy (MAE) of
a single molecule and its adsorption site \cite{SI}. 
The unit cell contains
an admixture of FCC, top and bridge adsorption sites. The single-molecule
MAE calculations yield a shift of $-0.47$ meV for molecules sitting on FCC and bridge
sites with respect to the gas-phase molecule, matching the pattern at larger bias.
{Figure}~\ref{figure2} (a)
is in good agreement with (d). On the top site, the
shift is $-0.67$ meV, explaining the experimental pattern at lower bias. 
Albeit small, the MAE dependence on adsorption site can be traced
back to the hybridization of vertical molecules with the substrate, see the supplementary information \cite{SI}.
Figure \ref{figure4} (a) sketches the ground-state spin texture of the resulting molecular adlayer.

We also evaluated the intermolecular magnetic interactions
using different non-collinear spin configurations, minimizing the energy
and fitting a generalized Heisenberg exchange tensor \cite{Choi2019,SI,Samir,Rebola}.
The interactions along the vertical dimer rows, red arrows in Fig. \ref{figure4} (a),
are ferromagnetic with isotropic exchange values of 1.26 meV
intradimer and 1.59 meV interdimer along the row. The horizontal
dimer rows are antiferromagnetic with isotropic exchange couplings
of 1.61 meV intradimer and 0.87 meV interdimer. 
Moreover, the interaction between vertical and horizontal molecules
amount to a small exchange coupling of 0.3 meV, while
the ferromagnetic interaction between vertical rows is a bigger exchange
coupling of 0.8 meV.
These facts draw a picture of  long-range interactions that facilitate
the collective reponse of the molecular adlayer to external perturbations.

\textcolor{black}{
The IETS lineshape is then a combination
of local molecular spin excitations and delocalized spinwave excitations.
Their combination} can be accounted for by using
Bloch spinwave theory \cite{Yosida}
and we evaluate the differential conductance as the convolution of BCS densities
of states in the presence of excitations \cite{SI}.
Figure  \ref{figure4} (b) shows the calculation of the $dI/dV$
for an isotropic 3-D RKKY exchange coupling and a local
 anisotropy $D= 3.7$ meV.
The experimental features are correctly described under the above assumptions, particularly
the extra broadening of the IETS peak compared to the QP one of Fig.\ref{figure3} (a).
The effect of the spinwaves is to broaden and shift the spectra of the local excitation, Fig.  \ref{figure4} (c).

%The emerging picture is of a complex non-collinear spin texture. Despite
%the decoupling effect of the superconducting gap, we do not expect
%strongly entangled spins on these systems because
%the large molecular MAE promotes Ising-like solutions \cite{Choi2019,Choi2017b}. 

%The direct electronic hybridization between vertical
%molecules is completely negligible due to the
%small lateral overlap of the $\pi$ frontier orbitals
%orginating in the cyclopentadienyl ligands. However, 
%we have seen that the overlap with
%the substrate is large.
%In these conditions, we should expect to find in-gap Yu-Shiba-Rusinov
%states \cite{Heinrich2018}
%induced by the vertical molecules.

\begin{figure}
\begin{center}
  \includegraphics[width=0.50\textwidth]{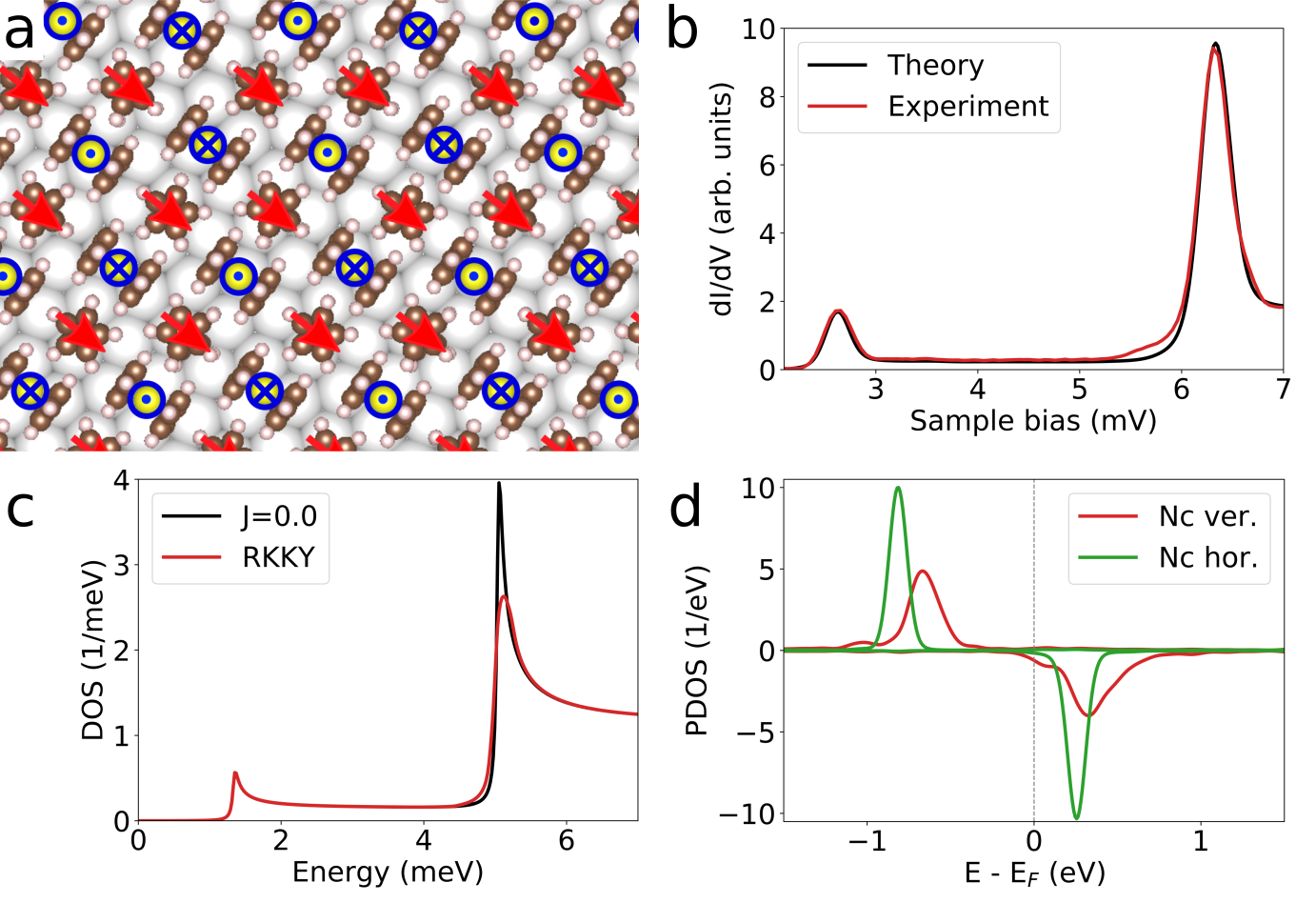}
\end{center}
  \caption{(a) Spin texture on the Nc \textit{paired} pattern.
\textcolor{black}{
The vertical-molecular  rows present ferromagnetic coupling (surface in-plane red arrows), 
while the horizontal molecules
have antiferromagnetic coupling (blue crosses and points representing inward/outward
arrows respectively). Due to the large axial magnetic
anisotropy, the spins are constrained to the plane perpendicular to the
molecular axis. 
(b) Simulated $dI/dV$ (black) compared
with the experimental one (red) using Bloch's spinwave theory and a molecule--molecule
3-D RKKY exchange coupling. (c) The red line is the electronic DOS used in the $dI/dV$
of (b) and the black line is the DOS for a zero intermolecular exchange coupling  ($J=0.0$). The
spinwaves shift the inelastic peak to higher energies and present a {broadened} DOS.
(d) DOS projected on the Nc $\pi$ frontier orbitals. The red curve
corresponds to vertical molecules, that hybridize with the substrate shifting to higher energies
the molecular orbitals and broadening them. The green curve is the DOS
for the horizontal molecules displaying a weak interaction with the substrate.} 
\label{figure4}}
\end{figure}

\textcolor{black}{
A complex non-collinear spin texture like the one above, on a superconducting
surface is prone to display Majorana states  \cite{Pientka2013,Pientka2015,Bernevig}.
With this aim, we study the presence of in-gap Yu-Shiba-Rusinov states \cite{Heinrich2018}
by first evaluating the electron-molecule exchange interaction and next the
value of the in-gap eigenenergy.}
Figure \ref{figure4} (d) shows the projected density of states (DOS) of
a Nc molecule on a bridge site. The symmetrically occupied/empty states about the
Fermi energy correspond to the majority/minority spins of the Nc $\pi$ frontier
orbitals. From the orbital splitting we can estimate the molecular
charging energy to be $U\approx 0.9$ eV. The peaks present a broadening $\Gamma \approx 0.2$ eV. Using the Schrieffer-Wolf transformation \cite{Yosida} we obtain that $\pi J_K\rho \approx 0.44$, where $J_K$ is the Kondo coupling and $\rho$ is the normal-state DOS at
the Fermi energy. The Yu-Shiba-Rusinov state energy for
this symmetrical case is $\epsilon_0=\Delta (1-\alpha^2)/(1+\alpha^2)\approx 0.8 \Delta$, where $\alpha=\pi J_K\rho S/2$, with $S=1$ for Nc. 
This type of system closely resembles the Shiba-chain limit
studied by Pientka and co-workers \cite{Pientka2013,Pientka2015}
and the dilute limit of 2D islands studied by Li \textit{et al.} \cite{Bernevig},
which can show topological phases characterized by large Chern numbers
and thus have several helical Majorana modes.  

A careful inspection of the superconducting gap around
the edge of the Nc islands permits us to conclude
that there is no such state and hence no
 topological phase \cite{Pientka2013,Pientka2015,Bernevig}.  
Probably due to the closeness of the in-gap states to the QP peaks.
Compressing the molecular layer \cite{Laetitia2018,Loth2018}, 
functionalizing the molecules or enhancing
the reactivity of the superconductor surface are strategies that
can approach in-gap state towards zero energy and
drive the Nc/Pb (111) system into the topological phase.

In summary, Nc molecules self-assemble aligning their molecular axis in
two different directions, creating molecular islands
with non-collinear spin structures. The role of the substrate is crucial
in creating the spin texture that translates into geometrical patterns
of the IETS signal as well as a broadened lineshape due to the presence
of local spin excitations and non-local spinwaves. The lack of direct
interaction between vertical molecules makes of this
system a first example of dilute non-collinear spin impurities that can
create topological superconductors characterized by large Chern numbers.
We believe that topological phases can be made accessible by small
modifications of the molecular islands.

These results show that molecular layers can create rich spin
textures through the interplay of different interactions.  This has
important implications on  superconductors because these spin textures
can create domains of different topological character, leading to new
material design.

%\acknowledgements
Financial support from the Spanish MICINN (project RTI2018-097895-B-C44), the
European H2020 FET open project MeMo  (grant no. 766864) and the French 
Agence Nationale de la Recherche (grants no. ANR-13-BS10-0016, ANR-11-LABX-0058 NIE and ANR-10-LABX-0026 CSC) are gratefully acknowledged.

\section*{Supplemental material: Molecular templates of spin textures on superconducting surfaces}

\section*{Density functional theory calculations}
Calculations have been performed in the framework of density functional theory as implemented in the VASP code \cite{Kresse1996b} using the projected augmented-wave (PAW) method \cite{Kresse1999}. Wave functions have been expanded using a plane wave basis set with an energy cut-off of 400 eV. We used  PBE as GGA-funcional \cite{PBE}. This functional was completed with dispersion corrections introduced through  the 
D3-BJ scheme \cite{D3BJa,D3BJb}. The Pb (111) surface was simulated using a 5 layers slab
with a vacuum region of 16~\AA. The three upper layers plus the molecules were relaxed
until forces were smaller that 0.02~eV/\AA. The calculations of isolated molecules and
dimers were performed for a  $6 \times 4 \sqrt{3}$ Pb (111) cell, using a k-point sampling
of $3 \times 3 \times 1$.

STM images have been simulated within the Tersoff and Hamann approximation \cite{tersoff_theory_1985} using the method described by Bocquet et al. \cite{bocquet_theory_2009} as implemented in STMpw \cite{lorente_stmpw_2019}.

\subsection*{Determination of the molecular adlayer}

\begin{figure*}
  \includegraphics[width=0.7\textwidth]{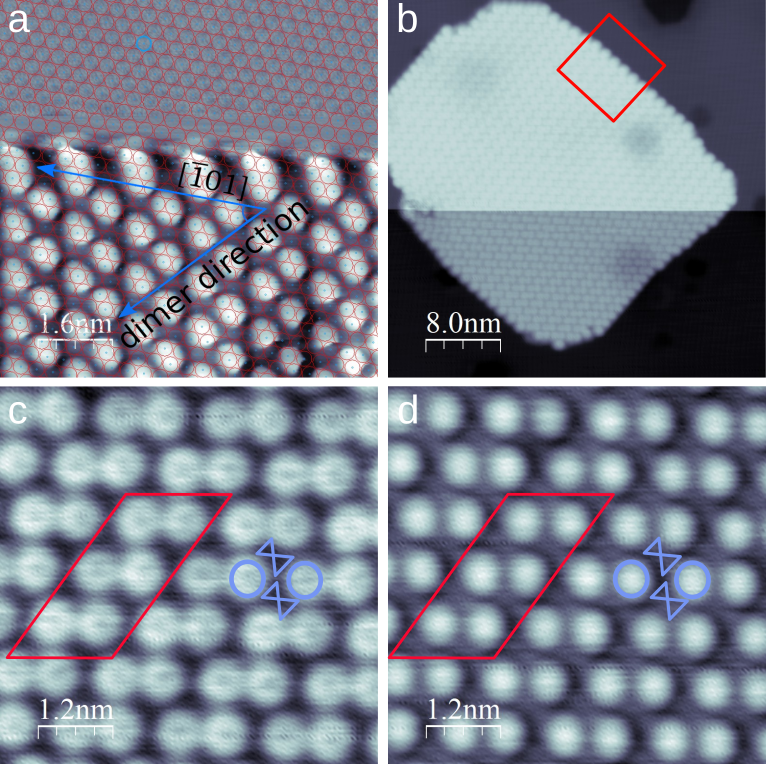}
  \caption{(a) STM image with surface atomic resolution
in the upper part of the image, the red circles indicate the hexagonal lattice on the Pb (111) surface. 
(b) Lower-current image, permitting to visualize the full Nc island. The red square marks the area of image (a). 
(c) Zoomed area of the molecular layer at 20 pA and $-20$ mV.
The blue 
schemes depict two vertical molecules (circles) and two horizontal ones (hour-glass schemes). (d) Same as (c) but with a bias of $-5$ mV.
\label{atomic}}
\end{figure*}

Atomic resolution on Pb (111) was obtained approaching the tip to the
substrate. Figure \ref{atomic} (a) shows atomic resolution close to
a Nc island. From the determination of the surface main direction,
we can determine the orientation of the island. We obtained an angle of
$47-48^\circ$ between the dimerization direction of the Nc layer and the
$[\Bar{1}01]$ direction of the (111) Pb crystal.  Figure \ref{atomic}
(c) and (d) correspond to constant current images of the molecular
layer at $-20$ mV and $-5$ mV respectively. At a bias of $-20$ mV the ring of the
molecule is most visible, at $-5$ mV a small corrugation can be observed at
the expected location of the horizontal Nc. The blue hour-glass (circle)
indicate the position of the horizontal (vertical) molecules. The unit
cell shown on Fig.~2 from the main text is also depicted here in red.

The structure of Fig.~2 of the main text for the linear arrangement
was obtained looking for a commensurate cell fulfilling (i) the above
orientation with respect to the surface, (ii) the dimer structure
observed in the topographic images, and (iii) the supercell emerging
from the dI/dV maps showing three dimers. Taking into account all these
conditions and patterns found in previous works \cite{Bachellier2016},
a 12-molecule supercell was proposed and optimized. The supercell
is determined by vectors $\begin{matrix}(5&-1)\end{matrix}$ and
$\begin{matrix}(3&7)\end{matrix}$ which form an angle of 53.9$^\circ$,
in good agreement with the $50\pm4^\circ$ experimental angle. The angle
with the $[\Bar{1}01]$ direction of the surface is 48.7$^\circ$, also
in good agreement with the experimental data.

The molecular adlayer was consequently shifted with respect to the
surface, so as the first molecule occupied a top site, then HCP and FCC
hollow sites, and a bridge site, originating 4 different configurations
for the surface structure.  The 4 structures were relaxed and their final
energies were within 20 meV/molecule, showing the small corrugation of
the molecule-surface potential energy surface.

We performed extra calculations of a single molecule and a single
dimer supported on the Pb (111) surface. From these calculations we
can glean more information about the interactions of the molecules
with the surface.  The adsorption energy is defined as the energy
needed to desorb the molecule. This can be obtained by subtracting
the energy of the molecule on the surface fully-relaxed system from
the free molecule energy plus the free-surface energy (fully relaxed).
The resulting positive number is the energy needed to go from
the adsorbed system to the desorbed one assuming there are no barriers
in-between these two asymptotic configurations.

The adsorption energy of a single vertical molecule is 0.668 eV.  From
here the vdW contribution is 0.435 eV leaving a chemical hybridization
of 0.233 eV. The dimer adsorbs with an energy of 0.682 eV/molecule,
the extra binding energy comes from the vdW interaction between the
two molecules. The adlayer shows an even larger adsorption energy per
molecule (0.939 eV/molecule) due to the extra stabilization furnished
by the intermolecular interactions. Indeed the cohesion energy of the
adlayer in the gas phase is 0.438 eV per molecule. As expected, it is
more difficult to desorb a single molecule from the molecular adlayer
on the surface than a single molecule on the surface.

\begin{figure*}
  \includegraphics[width=\textwidth]{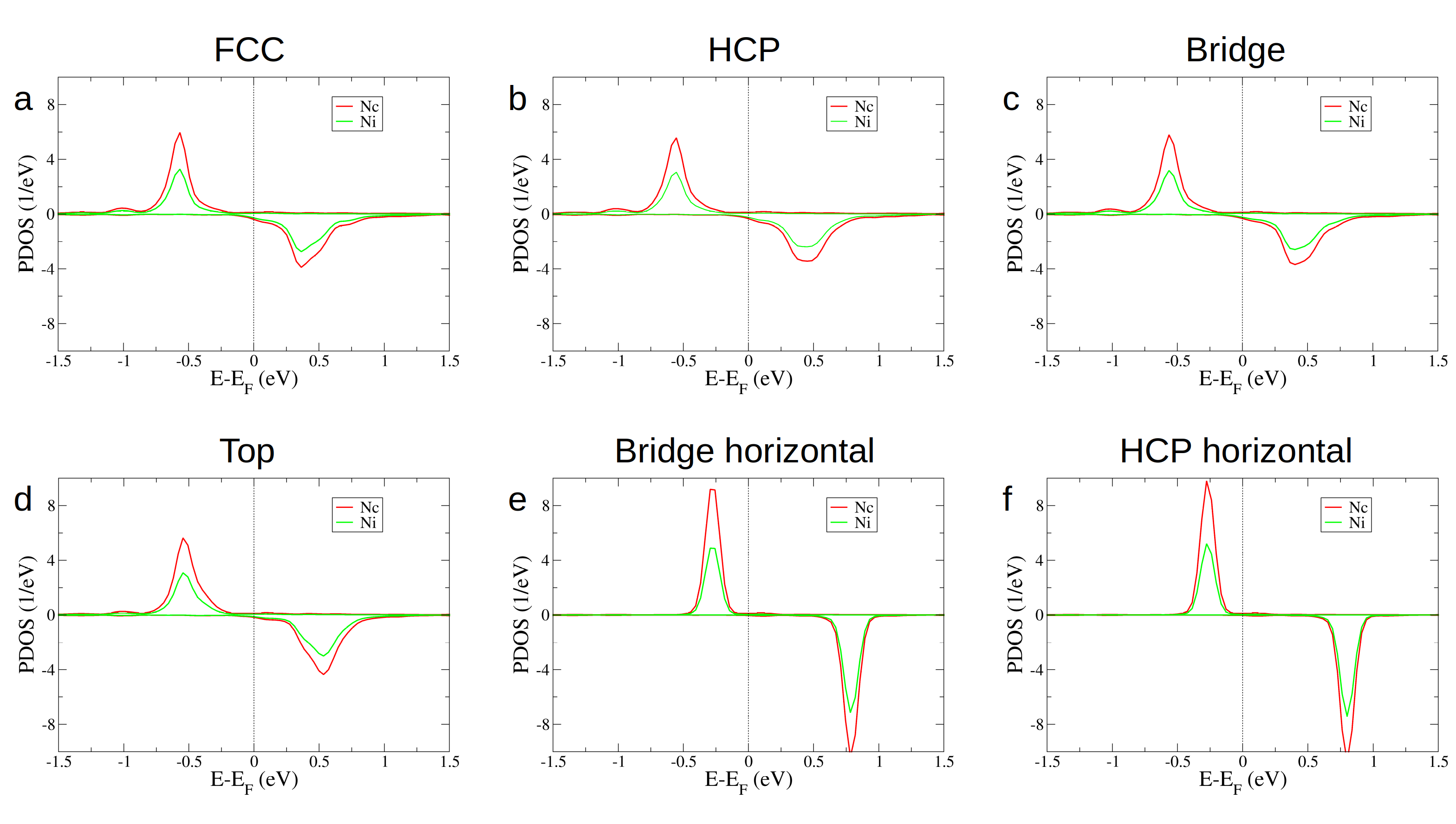}
  \caption{Projected densities of states (PDOS) of a single molecule on the surface at
different adsorption sites. (a-d) shows the PDOS at vertical molecules, while (e-f) represents the PDOS at horizontal ones. A Gaussian broadening with $\sigma=50$~meV has
been applied.
\label{pdos}}
\end{figure*}

Additional information can be extracted from the projected
density of states (PDOS) of the single molecules on the surface
(Fig.~\ref{pdos}). The comparison between the PDOS of vertical and
horizontal molecules shows the higher degree of hybridization of the
vertical molecules, where the $\pi$ orbitals are pointing towards the
surface. Among the vertical molecules the top case presents an upshift
of around 100~meV with respect to the other cases.

\subsection*{Linear and zig-zag arrangements}

\begin{figure*}
  \includegraphics[width=\textwidth]{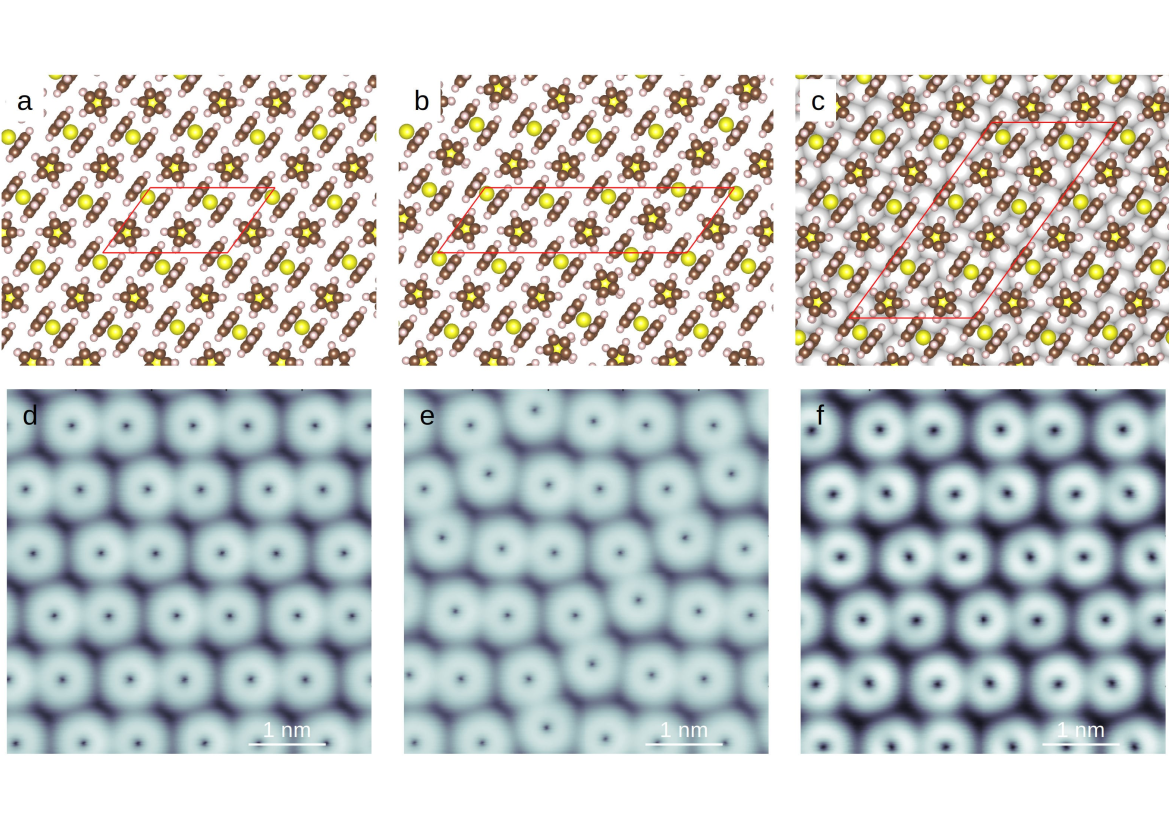}
  \caption{Unsupported structure of the linear (a) and zig-zag (b) arrangements of dimers.
The corresponding simulated STM images are shown in (d) and (e), respectively. (c) shows
the linear arrangement supported on Pb (111) with the unit cell derived from the dI/dV maps, 
with the corresponding simulated STM image shown in (f).
\label{stm}}
\end{figure*}

As explained in the main text, we find two arrangements after
deposition of the molecules: linear and zig-zag. We have studied
both of them in the unsupported case. Without considering the
surface the linear configuration can be described using a unit cell
containing two vertical and two horizontal molecules, as can be seen
in Fig.\ref{stm}a. In order to describe the zig-zag configuration the
unit cell has to be doubled to accommodate four vertical and four
horizontal molecules (Fig.\ref{stm}b). The corresponding simulated
STM images in Fig.\ref{stm}d and Fig.\ref{stm}e are in good agreement with the
experimental images, taking into account that we are not considering
the surface here. For comparison we have calculated the STM image of
the supported case in Fig.\ref{stm}f. Energetically linear and zig-zag
arrangements are close. We obtain a binding energy of 0.438~eV per
molecule for the linear case and 0.425~eV for the zig-zag one. Although
the calculations have been done in the gas-phase, the mainly vdW nature
of the interaction with the surface and the small corrugation of the
molecule-surface potential energy surface shown before make us believe
that both arrangements will be energetically close also on the surface,
in agreement with its coexistance in the experiment.

\subsection*{Single-molecule magnetic anisotropy}
Table \ref{MAE}
summarizes the main results concerning the calculations of magnetic anisotropy for single molecules. 
The anisotropy is calculated like in Ref. \cite{Ormaza2017a} and reproduces the overall values. 
The energy of the molecule is computed for a non-collinear spin configuration
with spin-orbit coupling. The energy is minimized self-consistently for two different directions
of the spin, one along the molecular axis and the other one perpendicular to it. Since
the molecular spin is $S=1$, the energy difference between both configurations readily
gives the value of $D$, as for example used in Eq. (\ref{HamSW}). The values are about a factor of two smaller than the experimental ones, as it is common of DFT
calculations with spin-orbit coupling \cite{Choi2017b,Panda,shick_2009} .
In this work, we are interested in the trend of the MAE with adsorption site, that
is more reliable than MAE absolute numbers.

Table \ref{MAE} shows that the MAE strongly depends on the adsorption site. 
This explains the distribution we find of the $dI/dV$ peaks, because they correlate the geometrical position of the molecule with the value of its MAE.
\begin{table*}
\begin{tabular}{llll}
\hline
Vertical molecule & & & \\
\hline
Site & magnetic moment ($\mu_B$) & Ads. energy (eV) & D (meV) \\
Free & 2.0 & -- & 1.77 \\
FCC & 1.77 & 0.67 & 1.30 \\
HCP & 1.77 & 0.65 & 1.50\\
Bridge & 1.78 & 0.65& 1.30\\ 
Top & 1.83 & 0.61 & 1.10 \\
\hline
Horizontal molecule & & & \\
\hline
Bridge & 1.95 & 0.54 & -1.72 \\
HCP & 1.95 & 0.52 & -1.71 \\
\end{tabular}
\caption{Calculated single-molecule adsorption and anisotropy energies
togehter with the molecular magnetic moment for single Nc molecules on Pb (111).
\label{MAE}}
\end{table*}

\subsection*{Exchange-coupling evaluation}

After minimizing the energy of different spin configurations with spin-orbit coupling, we
find that vertical dimers are ferromagnetically coupled in the supercell. Horizontal
molecules are antiferromagnetically coupled. The spins are at about 14$^\circ$ off
the vertical dimer rows and in-plane, while the horizontal dimers present out of plane spins. 
No degree of chirality is discernible from these calculations.

The exchange coupling is obtained by fitting a generalized Heisenberg interaction of the
form:
\begin{equation}
\hat{H}=\hat{H}_0 + \sum_{i,j} \vec{S}_i \cdot J_{i,j} \cdot \vec{S}_j.
\label{tensor}
\end{equation}
Here, $J_{i,j}$ is a tensor on the spatial coordinates of the spins $\vec{S}_i $
and $\vec{S}_j$. In the absence of
spin-orbit interaction the SU(2) symmetry is preserved and the tensor becomes
a scalar yielding the usual Heisenberg interaction. The spin-orbit coupling
breaks the spin symmetry and gives rise to anisotropic terms. Usually, the
tensor is then divided in three terms \cite{Choi2019,Samir}, an isotropic one
that is the Heisenberg exchange ($tr(J_{i,j})/3$) and the anisotropic one
that is the tensor without the trace term. The anisotropic tensor is further
divided in antisymmetic and symmetric tensors because the antisymmetric one 
exactly yields the Dzialoshinkii-Moriya interaction.

On Pb surfaces the spin-orbit coupling cannot be neglected. Even if the isotropic Heisenberg
interaction is the only one to be considered, the results greatly vary with and
without spin-orbit coupling.

To obtain the isotropic Heisenberg interaction, only two collinear configurations
are needed per interaction. This permits us to simplify the calculations. For the case of a single dimer,
we further use two chiral configurations to obtain the anisotropic
contributions to the tensor because the local longitudinal anisotropy of the Nc molecule
constrains the spin to the molecular plane. The exchange tensor of Eq. (\ref{tensor}) is
only a $2\times 2$ matrix in this case. The diagonal components are determined from
the collinear calculations and the off-diagonal ones from the chiral configurations.

%Shifted here from the main text

An isolated dimer (7-\AA\ distance between sites FCC and top)
shows an antiferromagnetic isotropic exchange of 0.99 meV,
and out-of-plane Dzialoshinskii-Moriya interaction $D_z=0.11$ meV and
an anisotropic symmetric exchange of $J_s=0.09$ meV.
The collective effect of the adlayer leads to a reversal of the isotropic
exchange interaction. Recent calculations \cite{Rebola} on Pb (110) show
that the three terms of the exchange tensor that we just mention
do present RKKY-like oscillations and can be of rather long range.
Despite the small contribution of the anisotropic terms of the exchange
tensor in our calculations, we cannot rule out a larger anisotropic
contribution leading to chiral solutions of the spin texture for
the adlayer.

\section*{Spin wave Bloch theory}

Bloch theory is a semiclassical approach (see for example the book
by Yosida \cite{Yosida}) derived from the electromagnetic solution
of an effective magnetic field obtained by fitting a mean-field approximation
to the following Hamiltonian:
\begin{equation}
\hat{H}=\sum_{i,j} J_{i,j} \vec{S}_i \cdot \vec{S}_j + \sum_i D_i S_{z,i}^2.
\label{HamSW}
\end{equation}
Where $\vec{S}_i$ is the spin on site $i$, $J_{i,j}$
is the isotropic exchange interaction between spin on sites $i$ and $j$, $D_i$
is the local longitudinal anisotropy on site $i$ and $S_{z,i}$
is the $z$ component of the spin on site $i$.
The assumption of the Bloch theory is a mean-field one where
the spins follow the electromagnetic field created by the
other spins and this gives rise to the Bloch equations.

For low-energy excitations, we can assume that spins  change
in only one unit. This is compatible with the spin-flips induced
by a tunneling electron. Looking at the small variations with
respect to the ground state (all spins aligned, ferromagnetic
case) in the electromagnetic equations, the Bloch
equations can be written like a harmonic oscillator of frequency
$\omega_q$, where we assume already a 2-D spin array and
wave vector $q$:
\begin{equation}
\hbar \omega_q =  (D  + J(0)- J(q)) S,
\label{freq}
\end{equation}
where the Fourier transform of the exchange coupling is defined as
\[J(q)=\sum_{\vec{R}_j} J_{j,0} e^{-i \vec{q} \cdot \vec{R}_j}\]
with $\vec{R}_j$ the lattice vector running over all the spin-array sites, $j$. 
These excitations are spinwaves that take into account
the energy to flip a spin in the presence of the
large molecular MAE, $D S_z^2$.

For the case of the Nc molecules, we take a simple RKKY interaction:
\begin{equation}
J(q)=\sum_j J_0 cos(\vec{q} \cdot \vec{R}_j) cos (2 k_F R_j)/R_j^3,
\end{equation}
where $R_j$ is the modulus of the lattice vector $\vec{R}_j$.
Both $k_F$ 
and $J_0$ are optimized to approximate the DFT values for the intra- and inter-
dimer ferromagnetic exchange interactions.
Neglecting the RKKY interaction and just taking a nearest-neighborg
$J$ leads to van Hove-like singularities in the excitation spectra
that are not realistic.

\section*{Simulation of differential conductance}

Using the same approach of Refs. \cite{Ruby2015a,Choi2017}, we assume that the current is proportional to the convolution of the tip's and sample's densities of states, $\rho_T (\omega)$ and $\rho_S (\omega)$, respectively. Hence, the differential conductance is given by
\begin{eqnarray}
\frac{\partial I}{\partial V} &\propto & \int \rho_S (\omega) \frac{\partial\rho_T (\omega-eV)}{\partial V} [ f (\omega-eV,T)
-f (\omega,T) ] d \omega \nonumber \\
&+& \int \rho_S (\omega) \rho_T (\omega-eV)  \frac{\partial f (\omega-eV,T)}{\partial V} d \omega
\label{cond}
\end{eqnarray}

For the tip, we assume a simple BCS density of states (DOS)
with a phenomenological broadening Dynes parameter $\gamma$ to broaden
the quasiparticle singularities:
\begin{equation}
\rho_{BCS}(\omega) \propto \text{sgn}(\omega)\text{Re}\left(\frac{\omega + i \gamma}{\sqrt{(\omega+i \gamma)^2 - \Delta^{2}}}\right).
\label{bcs}
\end{equation}

For the substrate, $\rho_S (\omega)$ is computed using the spinwave excitations, Eq. (\ref{freq}). The electronic density of states, $\rho_S (\omega)$, is the
imaginary part of the 
propagator of an electron in the presence of spin excitations. The electron yields $\hbar \omega_q$ every time it excites a spinwave. In order for this process to be possible, it
needs to end up in an empty state of 
the BCS density of state. As a consequence, the
electronic density of states in the presence of spin excitations is approximately:
\begin{equation}
\rho_S (\omega)= A \rho_{BCS}(\omega) + \sum_q \rho_{BCS}(\omega-\omega_q).
\label{dos}
\end{equation}
The first term is the elastic contribution, or the density of states
without excitations. It is weighed by a parameter $A$ to fit the ratio
of elastic to inelastic contribution.
The second term yields the contribution of each spinwave to the electronic
density of states. Since the spinwaves are equivalent spin-flip excitations,
we assume they enter with the same weight.

The calculations are performed using the experimental temperature, 2.5 K, and a gaussian broadening of 180 $\mu$eV due to RF-noise and lock-in amplification in order to reproduce the experimental QP conductance. The same parameters
where later used for the IETS peak.

\bibliography{References}

\end{document}